# Relaying Topological Interface States for Negative Refraction of Bulk Waves

Hongchen Chu[1†], Ze-Guo Chen[2†,*], Yun Lai[1*], Guancong Ma[2*]

[1]*National Laboratory of Solid State Microstructures, School of Physics and Collaborative Innovation Center of Advanced Microstructures, Nanjing University, Nanjing, 210093 China*

[2]*Department of Physics, Hong Kong Baptist University, Kowloon Tong, Hong Kong, China*

[†]These authors contributed equally to this work.

[*] Correspondence to: zgchen@hkbu.edu.hk, laiyun@nju.edu.cn, phgcma@hkbu.edu.hk.

Topological notions in physics have become a powerful perspective that leads to the discoveries of topological interface states (TISs). In this work, we present a scheme to achieve negative refraction by leveraging the properties of TISs in a valley photonic crystal (VPC). Due to the chiral characteristics, one type of the TISs deterministically possesses negative dispersion relation, which can cause an obliquely incident wave to undergo a negative lateral shift. By stacking multiple VPC interfaces, the TIS-induced lateral shifts can relay in the transmitted wave towards the far side of incidence. The resultant outgoing wave appears to have undergone negative refraction. This finding is verified in microwave experiments. Our scheme opens new application scenarios for topological systems in bulk wave manipulations.

Since its first introduction to physics in 1980s, the concept of topology has drastically changed our fundamental understanding of matters (*1*, *2*). The new insights brought about by topological notations have led to the discovery of a plethora of exotic phenomena. But perhaps none is more impactful than the deterministic prediction of interface states localized between two topologically distinct matters (*3*, *4*). These topological interface states (TISs) have the fascinating properties of being robust against disorder and sometimes immune to backscattering, making them an ideal candidate for transferring information or energy. In recent years, the exploration of topological notions in photonics and electromagnetism (*5-7*), and acoustics (*8*, *9*), have also unshackled our traditional understandings of these classical waves and led to exciting new developments. The localized characteristics of topologically protected states make them a natural candidate for manipulating confined waves, e.g., at boundaries or interfaces. It is therefore not surprising that the potential of TISs for bulk wave steering applications has largely been ignored.

In parallel with the development of topological notions in classical wave systems, bulk waves steering has witnessed great advancements in the past decades, thanks to the developments in wave-functional materials such as metamaterials. Among a kaleidoscopic collection of novel and exotic wave phenomena, negative refraction perhaps can take the crown of being the most impactful one. So far, negative refraction was obtained by negative effective index or suitable dispersion in metamaterials and wave crystals (*10-25*), phase gradience in metasurfaces (*26-30*), metasurfaces with parity-time symmetry (*31-34*), etc. However, none of the existing approaches anticipated that topological notions can play a role in achieving negative refraction. Here, we report a topological route to deterministically achieve negative refraction of bulk waves by leveraging on the properties of TISs. Our approach is based on the TIS existing at zigzag interfaces of two-dimensional (2D) valley-Hall photonic crystal (VPC) (*35-41*). The chirality of the TISs is uniquely associated with the cleavage at the interface. We show that with a specific cleavage at the interface, the TIS must possess a negative dispersion. As a result, an incident wave would experience a position shift negative to the wave vector component along the interface. By stacking such VPC interfaces and optimizing the coupling between the neighboring layers, interestingly, multiple TISs can relay the incident wave towards the direction of negative refraction. In microwave experiments, we observed pronounced negative refraction for the beam exiting at the far side of the VPC superlattice. Besides negative refraction, positive refraction can also be produced by using a different type of interface.

We begin by considering a 2D VPC consisting of a honeycomb lattice of dielectric cylinders with relative permittivity $\varepsilon_r = 12.97$ and radii $r = 0.3a/\sqrt{3}$ in air, where $a$ is the lattice constant. The band structure for $[\vec{E} = (0,0, E_z)]$ is plotted in Fig. 1(b) as black dots, wherein Dirac cones are clearly seen at K(K'). Then, we break the inversion symmetry in the unit cell by tuning the radii of the cylinders to $r_1 = 0.24a/\sqrt{3}$ and $r_2 = 0.36a/\sqrt{3}$ (red and blue dots in Fig. 1(a), respectively). This lifts the degeneracies of the Dirac cones and creates two "valleys" at K(K'), as shown in the band structure in Fig. 1(b) (red dots), each possessing a local distribution of nonzero

Berry curvature. As a result, integration of the Berry curvature near each valley gives rise to nonzero "valley Chern numbers" (*42*, *43*), which protect the existence of chiral TISs.

We utilize the above method to produce two types of unit cells, denoted type-A and B, which are mirror-symmetric counterparts of each other. We construct a supercell strip that has one layer of type-B unit cells sandwiched by two layers of type-A unit cells, as shown in Fig. 1(a). Each layer has 20 unit cells. The supercell strip thus has two types of zigzag interfaces. Type-I interface, marked by the red dashed line (Fig. 1(a)), is joined by two smaller cylinders, whereas type-II interface is joined by two larger cylinders (blue dashed line in Fig. 1(a)). The strip's band structure is plotted in Fig. 1(c). A bulk bandgap is seen at 0.26 – 0.32 (normalized frequency). In the gap, TISs are found, which are plotted in red and blue, corresponding to the interface types at which they localize. These TISs have opposite chirality near the two distinct K(K') points. Importantly, the TIS at the type-I interface (red in Fig. 1(c)) has a negative dispersion for $|k_x| > 0.4$. This characteristic plays a crucial role in achieving negative refraction.

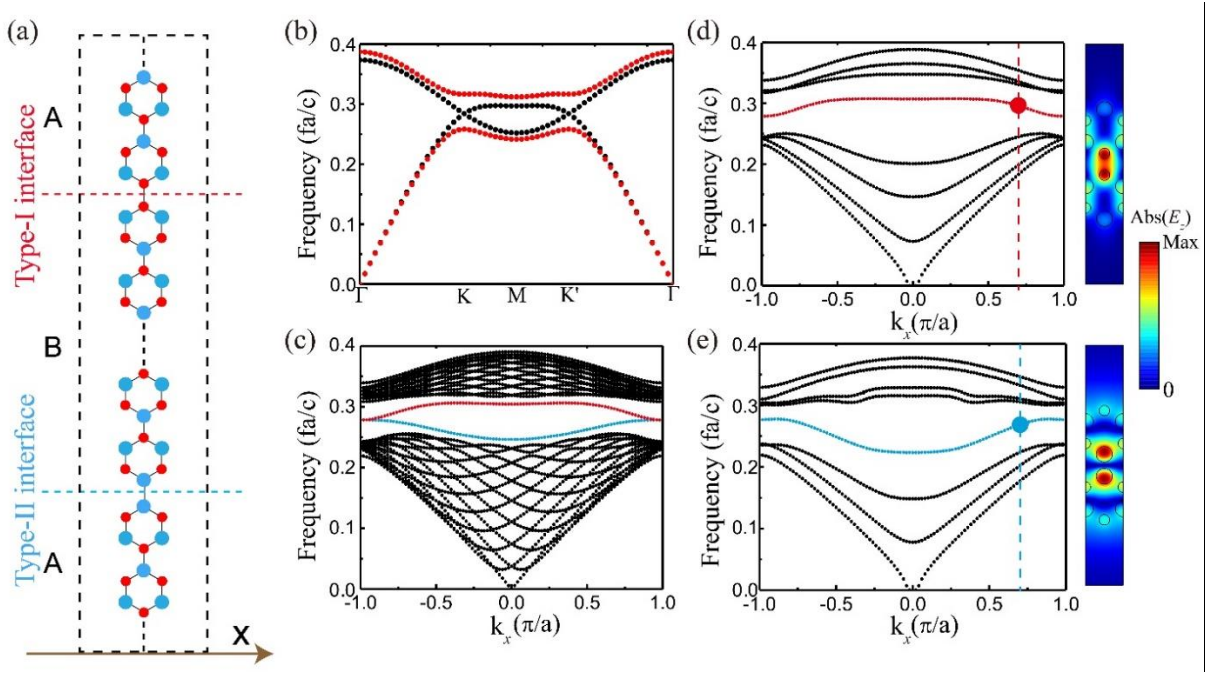

Fig. 1 (**a**) Schematic of a VPC strip containing two types of interface labeled by red and blue dashed lines. (**b**) The bulk band structure of the VPCs, black (red) dots represent the band structure without (with) radius modulations of the cylinders. (**c**) The projected band structure shows TISs with different chirality, the red (blue) dots indicate the interface state at the interface marked in (a). (**d**, **e**) The interface state behaves well-

localized even when the system is only composed of two layers of VPCs. The right inset shows the field distributions where color represents out-of-plane electric field amplitude.

Naturally, when the strip has only one type of interface, the corresponding TIS has deterministic dispersion. Such a characteristic has been leveraged to achieve a one-way interface transport of waves analogous to the valley-Hall effect (*35*). Surprisingly, such TISs persist even when the strip contains only two unit cells (one of each type), as shown in the band structures in Fig. 1(d, e). The TIS field profiles show strongly localized distributions at the interface with an evanescent characteristic seen in the *y*-direction. Such field profiles indicate a large real wavevector parallel to the interface and an imaginary wavevector in the normal direction.

We next examine the transmission of a wave incident on such a strip at the frequencies of the TISs. Consider a plane wave incident at an angle $\theta$, as schematically shown in Figs. 2(a) and 2(d), we compute the transmission coefficients $t(f,\theta)$ for the frequencies inside the bandgap and the incident angles $0 < \theta < 60°$. The results for both types of interfaces are respectively shown in Fig. 2 (b) and (e). Large transmission is identified at specific frequencies and angles. By considering the conservation of momentum (wavevector) along the interface direction, we retrieved the dispersion relations from the transmission spectra (*37*), which aligns excellently with the TIS bands, as shown in Figs. 2(c) and 2(f). This evidence proves that large transmission is possible by relaying the TISs, despite their individual evanescent characteristic. Indeed, the large transmission can be viewed as a generalization of extraordinary optical transmission (*44-46*), in which the contribution from surface plasmon polariton is replaced by the TISs.

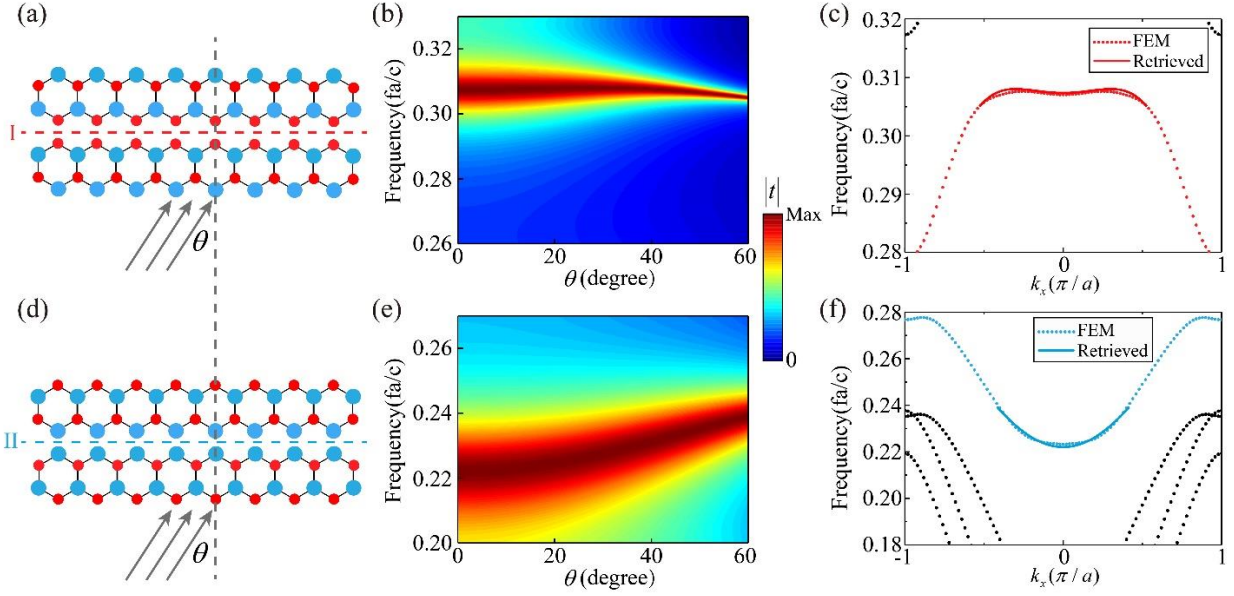

Fig. 2 Schematic drawing of a plane wave incident on a system with a type-I (**a**)/ type-II (**d**) interface. Both systems have only two unit cells. (**b**, **e**) The transmission spectra as functions of incidence angle $\theta$ and normalized frequency $fa/c$. The color represents the transmission amplitude. (**c**, **f**) The TIS dispersions retrieved from the transmission spectra (red and blue dots) show excellent agreement with the band structures calculated by FEM (red and blue solid curves).

   The chiral characteristic of the TISs at different interfaces enables a novel way to manipulate bulk wave transmission. With a type-I interface (Fig. 2(a)), the TIS has negative dispersion in $k_x$. Such a TIS would transfer wave energy toward the negative direction of the wave vector along the interface. As a result, when an incident wave couples to this TIS, the wave gains a negative shift parallel to the interface and the beam exiting from the far side would appear to have undergone negative refraction. However, for the type-II interface, the incident wave would gain a positive shift parallel to the interface, appearing to undergo positive refraction. We numerically verify these findings by FEM simulations. We employ a Gaussian beam incident at an angle of 45 degrees, as shown in Fig. 3(a) and (c). At a normalized frequency of $\frac{fa}{c} = 0.307$, negative refraction behavior is observed for the type-I interface, as the center of the outgoing beam is shifted to the left side of the normal (Fig. 3(b)). In contrast, positive refraction is seen for the type-II interface at a frequency of $\frac{fa}{c} = 0.233$, as shown in Fig. 3(d). The difference in frequency is due to the spectral mismatch of the TISs at different interfaces (Figs. 1(d, e)). We note that there is a certain amount of reflection for both cases, which is due to the fact that the incident Gauss beam carries additional wavevectors and impedance mismatch.

By stacking multiple type-I (type-II) VPC interfaces together, we obtain a superlattice as shown in Figs. 3(e, g). Although both type-I and type-II interfaces exist in the superlattice, the two types of TISs are mismatched in their frequency. By choosing the frequency of the incident wave, we can selectively excite only one type of TIS and suppress the other type. In this way, the oblique incident wave is relayed across the superlattice by only one type of TISs, resulting in a pronounced negative or positive shift in the outgoing beam, which depends on the dispersion of the chosen TISs. We take four layers of type-I (type-II) VPC interfaces as an example and calculate the field distribution by FEM simulations. Figures 3(f, h) show the calculated field distributions under the illumination of a Gaussian beam at an angle of 45 degrees. The outgoing beam is significantly shifted to the left (right) side of the surface normal, giving rise to pronounced negative (positive) refraction effect.

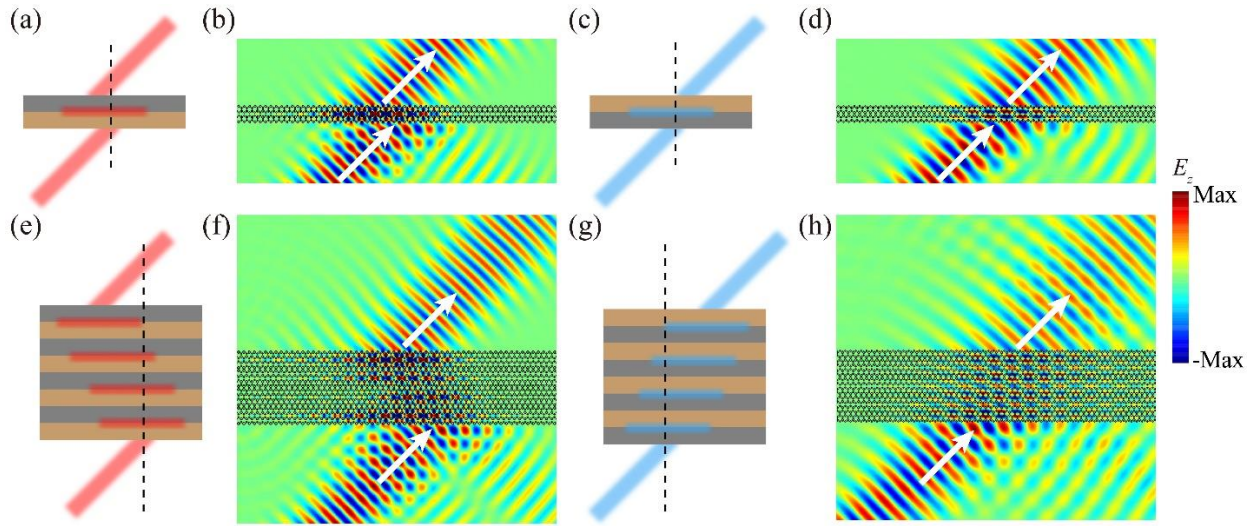

Fig. 3 Schematic of a two-layer system supporting transmission extreme where the transmitted wave performs as negative (**a**) or positive (**c**) refracted, depending on the type of the excited TIS. (**b**) and (**d**) FEM simulations show negative and positive refraction. The length of the system along the interface direction is $60a$. The white arrows indicate the direction of energy flow. Schematic of a multiple-layer system supporting negative refraction (**e**) and positive refraction (**g**). (**f**, **h**) FEM simulations verification.

Finally, we demonstrate the TIS-induced positive and negative refraction by proof-of-principle microwave experiments. The VPCs consist of a honeycomb lattice of alumina ceramics ($\varepsilon_r = 12.97$) cylinders with a height $h = 8$ mm and $a$ is 9.21 mm. The simulated electric field distributions $E_z$ for VPC superlattices with four layers of type-I and type-II interfaces are respectively shown in Fig. 4(a) and 4(d), where negative refraction and positive refraction are both clearly observed. The VPC is assembled in the *xy* plane inside a parallel-plate waveguide composed of two flat aluminum plates, as shown in Fig. 4(b), where the upper plate is not shown here. A standard X-band waveguide with absorbing materials is adopted to launch a beam with a finite width. To facilitate field scanning, the separation between the two aluminum plates is 8.7mm, which is slightly larger than the height of the VPCs (8mm) so that the lower plate can move by mounting it on a translational stage. The electric field is measured by an antenna fixed in a hole in the upper aluminum plate. The measured electric field distributions of VPC with four-layer type-I interfaces at frequency 11.32 GHz is shown in Fig. 4(c). It is clearly seen that the transmitted wave is shifted to the left side of the surface normal, which is consistent with the simulation results obtained by FEM in Fig. 4(a). Specifically, in Fig. 4(c), we see the measured fields inside the VPC have a layered pattern, which implies the excitation and relay of the TISs. These results prove our prediction of negative refraction by relaying TISs. On the other hand, Fig. 4(e) shows the measured electric field distributions at frequency 8.80 GHz, in which positive refraction is observed to be in excellent agreement with the simulation results shown in Fig. 4(c). Slight deviations of the working frequency between the experimental and simulation results are attributed to the air gap between the VPC and the upper plate of the parallel waveguide.

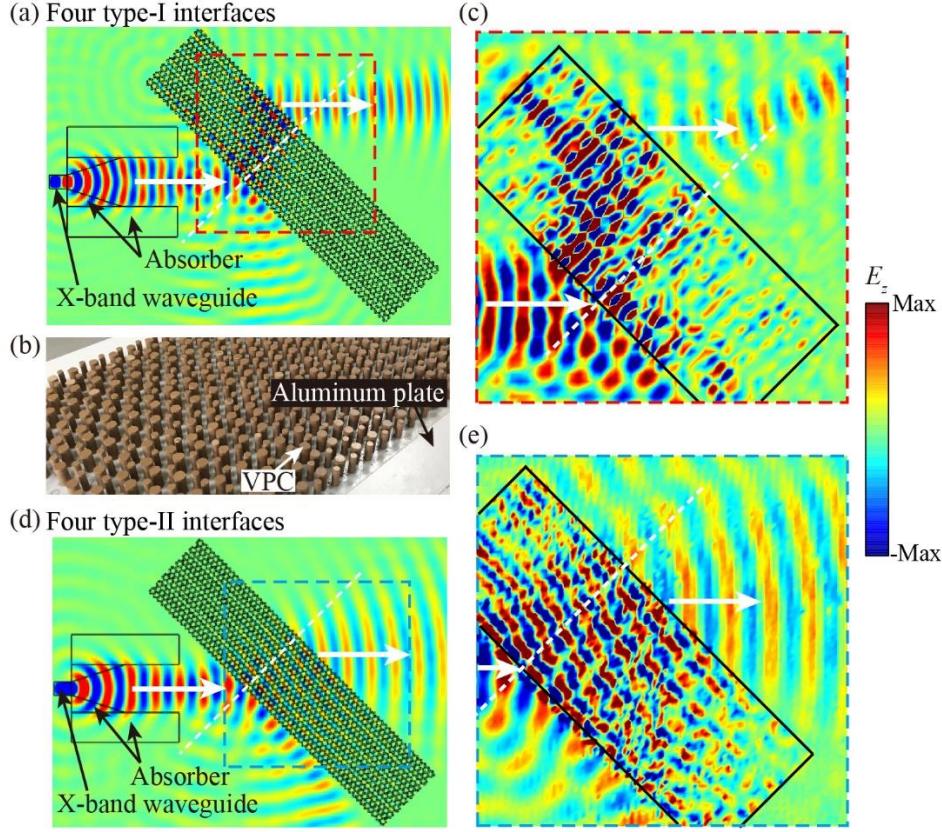

Fig. 4 Experimental demonstration of negative and positive refraction by relaying TISs. (**a**, **d**) Simulated field distributions of VPCs with four type-I (a) and type-II (d) interfaces. Dashed boxes depict the measured region in experiments. The incident beam is generated by a standard X-band waveguide along with absorber materials. (**b**) The photo of the fabricated VPC superlattice consisting of alumina ceramics cylinders. Measured field distributions of negative refraction through VPC with four type-I interfaces (**c**) and positive refraction through VPC with four type-II interfaces (**e**).

In conclusion, we proposed and demonstrated a novel strategy by stacking VPC interfaces, each sustaining a TIS to realize the negative refraction between the VPC superlattice and the background medium. Strikingly, although our mechanism is underpinned by TISs with negative dispersion, the corollary outcome is a bulk-wave steering effect in the superlattice. Therefore, our results hint that the functionality of TISs can go beyond edge/interface-wave manipulation. Meanwhile, it is worth noting that the VPC superlattice can also be analyzed by using the equifrequency contour of its bulk bands, which also indicates negative dispersion. With the powerful tool of topological analysis shown here, such negative bulk bands are revealed to have a topological origin. Such a topological origin dictates that the VPC interface must support two types of TISs with opposite chirality, and TISs with negative dispersions are guaranteed to exist. Consequently, our scheme is deterministic and robust against small perturbations in system parameters. Our design is thus fundamentally different from existing routes to negative refraction in metamaterials and metasurfaces requiring delicate design of effective indices or phase gradients. Our work also differs from the previous fermi-arc induced negative refraction, which guides waves along different surfaces of a 3D system (*47*, *48*). Since the valley Hall system must support two types of TISs with opposite chirality, TISs with negative dispersions are guaranteed to exist. Such a topological origin indicates that our scheme is deterministic and robust against small perturbations in system parameters. Lastly, the relaying mechanism of TIS is not restricted to valley Hall systems. It will also exist in other types of topological phases such as the Chern insulator model (*49*, *50*), 2D Su-Schrieffer-Heeger model (*51*, *52*), quantum spin Hall systems (*53*), and so on. The mechanism can also be adapted for other wave systems such as acoustic waves, elastic waves, and optics.

**Acknowledgments**

Z.-G. C. thanks Xiao Hu and Xiao-Chen Sun for helpful discussions. This work is supported by Hong Kong Research Grants Council (GRF 12302420, 12300419, ECS 22302718, CRF C6013-18G), National Science Foundation of China Excellent Young Scientist Scheme (Hong Kong & Macao) (#11922416) and Youth Program (#11802256), and Hong Kong Baptist University (RC-SGT2/18-19/SCI/006). Y. L. also acknowledges financial support from the National Key R&D Program of China (2020YFA0211300), the National Natural Science Foundation of China (Grants #11974176, #61671314).